\documentclass[pre,twocolumn,showpacs,superscriptaddress] {revtex4}
\usepackage{graphics}
\newcommand{\cu}
{\affiliation{Department of Physics, University of Calcutta,
92 Acharya Prafulla Chandra Road, Kolkata 700009, India}}

\begin{document}
  
 \title
{Nonconservative kinetic exchange model of opinion dynamics with  randomness and bounded confidence}

\author	{  Parongama Sen }
\email[Email: ]{psphy@caluniv.ac.in}
% \email{psphy@caluniv.ac.in}
\cu
\begin{abstract}

The concept of a bounded confidence level is incorporated in a  
nonconservative kinetic exchange model of opinion dynamics model where opinions have continuous values $\in [-1,1]$. The characteristics of the 
unrestricted model, which has one parameter $\lambda$ representing conviction, undergo drastic changes with the introduction of bounded confidence
parametrised by $\delta$. 
Three distinct regions are identified in the phase diagram in the $\delta-\lambda$ plane and the 
evidences of a first order phase transition for $\delta \geq 0.3$ are presented. A neutral state with all opinions equal to zero
occurs for $\lambda \leq \lambda_{c_1} \simeq 2/3$, independent of $\delta$, while for  
$\lambda_{c_1} \leq \lambda \leq  \lambda_{c_2}(\delta )$, an ordered region is seen to exist
where opinions of only one sign prevail.
At $\lambda_{c_2}(\delta )$, a transition to a disordered state is observed,
where individual opinions of both signs coexist and move closer to  the extreme values ($\pm 1$) as $\lambda$ is increased. 
For confidence level $\delta < 0.3$, the ordered phase exists for a narrow range of $\lambda$ only.
The line $\delta = 0$ is apparently a line of discontinuity and this limit is discussed in some detail.

\end{abstract}

\maketitle
%\begin{multicols}{2}

\section{Introduction}

Many social phenomena can be  modelled and analysed using the methods of statistical physics. 
Opinion dynamics is one such phenomenon where the microscopic
interactions taking place between individuals can lead to cooperative 
behaviour of the society, like the emergence  of a consensus or agreement. 
Recently, several opinion dynamics models with continuous opinion have been proposed 
in which the opinions are updated after a pair of individuals interact and exchange opinion.  
Here the  interaction scheme is similar to  a   
  kinetic theory of gas, where one assumes that gas molecules can collide and consequently
exchange  energy and momentum. 
Two basic  opinion dynamics models  having such kinetic exchange scheme are due to 
Deffuant et al \cite{deff} and Lallouache et al \cite{bkcop}, 
which  involve completely different concepts.
 In this paper, we attempt to combine the concepts used in these two basic models in a single model.
In section II, we briefly review these two models. In section III, the proposed model is described. 
The results are presented  and discussed in section IV. Summary and conclusions are given in the last section.

\section{A brief review of two kinetic exchange models}
 
Deffuant et al \cite{deff} introduced a simple model (DNAW model hereafter) in which opinion exchanges 
between two agents 
take place only when the  
difference in the original opinions is less than or equal to a preassigned quantity $\delta$.  
If $o_i(t)$ is the opinion of the $i$th agent interacting with the $j$th agent
at time $t$ (with $|o_i-o_j|\leq \delta$), then in this model the opinions evolve according to:
\begin{eqnarray}
\label{eq:deff}
o_i (t+1) &=&  o_i(t) + \gamma  (o_j(t)-o_i(t))  \nonumber \\
o_j (t+1) &=&  o_j(t) + \gamma  (o_i(t)-o_j(t)).
\end{eqnarray}
Here $\gamma$ is a constant ($0 \leq \gamma \leq 0.5$) called the convergence parameter and 
 $o_i$ lies in the interval [0,1].
The dynamics is such that the opinions tend to come closer after interaction.
Hence as the dynamics proceeds, convergence to a finite number of opinions
happens; opinions cluster around a few values and individuals belonging to different clusters no longer interact. The
initial distribution of the individual opinions is uniform and therefore symmetric. This 
symmetry is broken as the time evolved  distribution has a multipeaked delta
function form.  When there is only one 
peak in the final distribution, it is said to be a case of {\it consensus}, two peaks imply {\it polarisation} and the existence of a larger (finite) number of  peaks signifies {\it fragmentation} in the society.

 The model is conservative as total opinion
is conserved in each interaction. 
%As the original opinions are distributed uniformly, 
%symmetry breaking is said to take place as  the distribution of the  final opinions 
% cluster  around one or more discrete values.
 Obviously, in this conservative model, consensus would imply that opinions converge to the value  1/2.
Several models  have been formulated  incorporating the 
idea of bounded confidence later \cite{rmp,fortunato,hk} and  
a general form of kinetic exchange type model proposed in \cite{toscani}.
% A general class of models were introduced in \cite{toscani}
%where both apart from these exchanges, diffusion was also allowed. 

 More recently, a model in which kinetic exchanges take place with randomness, and where there is no conservation,   
has been introduced by Lallouache et al \cite{bkcop} (LCCC model hereafter). 
Any two agents can interact in this unrestricted model.
The opinion evolution here follows the rule:
\begin{eqnarray}
\label{eq:lccc}
 o_i(t+1) &=& \lambda[o_i(t) + \epsilon o_j(t)] \nonumber \\
 o_j(t+1) &=& \lambda[o_j(t) + \epsilon^\prime o_i(t)];
%\label{lccc-eq}
\end{eqnarray}
 where $\epsilon$, $\epsilon^\prime$ are drawn randomly from uniform
distributions in $[0,1]$.
In this model  $\lambda$ is a parameter
which is interpreted as `conviction'.
The opinions are bounded, i.e., $-1 \le o_i(t) \le 1$; in case $o_i$ exceeds 1 or becomes less than
$-1$, it is set to $ 1$ and $-1$ in the respective cases.

It is possible to rescale the opinions in the DNAW model so that they  lie in the interval $[-1,1]$.
Continuous opinions are relevant in cases like supporting a issue, rating a movie etc. Thus setting the 
interval as $[-1,1]$ appears to be more  
 meaningful  
since   a positive (negative) value of the opinion will mean liking (disliking) the motion.
The magnitude of the opinion would then simply correspond to  the amount of liking or disliking. For the rest of the paper, we thus consider  opinions $ \in [-1:1]$.

 As there is no conservation in the LCCC model, the average opinion given by  $m= |\sum_i o_i |/N$
for a population of $N$ agents,  evolves in time and  $m$ 
   can play the role of an order parameter.  It is, in fact,  analogous to magnetisation in magnetic systems. One can say that  there is order/disorder when
$m(t \to \infty)$  converges to a nonzero/zero value.
%Ordered and disordered states in this model are    analogous to ferromagnetic and paramagnetic states in 
%magnetic systems. 
The model shows a unique
behaviour: below a  critical value of $\lambda \simeq 2/3$, all opinions identically turn out to be zero while
above it, there is a nonzero value of the average opinion.
Thus for $\lambda> 2/3$, an ordered phase exists.
  Interestingly, the 
opinions in the ordered phase  have either all positive or all negative values.
Generalisation and variations of the LCCC model have been considered in some subsequent works \cite{ps,soumya,soumya2}.

Symmetry breaking has different connotations in the conserved DNAW model 
and the nonconserved  
LCCC  model. In the former, if opinions are initially    
in the interval $[-1,1]$, 
a consensus implies  convergence of all opinions to zero value and this is 
regarded as symmetry breaking as mentioned earlier.
In LCCC, the identical state of all zeros is also obtained below   $\lambda \simeq 2/3$   
even without putting any restriction on the interactions. 
However, this state  has been  interpreted as a symmetric state \cite{bkcop}. This 
is following the idea
that as  $m=0$ here, it is like a paramagnetic state  (which is a symmetric state in magnetic systems).
But obviously, this is a very special paramagnetic state which also has zero fluctuation. 

\section{The model}

Having conservation in the opinions  is rather unrealistic but the
concept of having a bounded confidence level is relevant in many cases.
On the other hand, it is true that people may have a conviction level in their
and in others' opinions. 
We thus combine the LCCC model and the DNAW model by putting the restriction of the bounded confidence in the former. 
It has to be realised that the conviction and bounded confidence are independent concepts 
and the parameters associated with these may be regarded as independent. Putting  a bounded 
confidence may be  applicable to cases in which there are  groups of people (e.g., members of a  political 
party) who share similar opinion and interact within themselves primarily. On the other hand, conviction
is associated with individual behaviour.
It is known that in models with bounded confidence, for small confidence level, the society is fragmented 
because of lack of communication.   A  large value of conviction 
is not expected to help reach consensus when the confidence level is low 
when we combine the two concepts,  as the groups stick to their 
original opinions  in this case. On the other hand, even when confidence level 
is large, very large values of conviction can  make the society polarised as 
both positive and negative opinions can prevail simultaneously. 

In the model proposed in the present paper, we follow eq (\ref{eq:lccc}) for the evolution 
of opinions but put the restriction that  
%Since opinions range from $+1 $ to $-1$, we define $0 \leq \delta \leq 1$
 agents interact only when $|o_i - o_j| \leq 2\delta$. $\delta$ is once again the  parameter 
representing the confidence level and can vary from zero to 1.   

We therefore have two parameters in the model, $\lambda$ and $\delta$. $\delta =1$ is identical to the original
LCCC model. $\delta = 0$ is an interesting limit. Here, agents interact only when their opinions are exactly same.
%Since in a continuous distribution of opinions, these cases will be extremely rare, we expect the system to be noninteracting in general
%and opinions will remain frozen, at least in a finite system. 
We will discuss this limit in greater detail later.

The order parameter in this model is the one defined for the LCCC model, i.e., $m = |\sum_i o_i |/N$.
To avoid confusion, 
    we adopt the following terminology for the present nonconserved model: when   $m=0$  and also the  
individual  opinion distribution is a delta function peaked at zero, we call  
it a {\it neutral} state.
If it is not a delta function at zero but $m =0$, 
the state  will be termed  
 a {\it disordered} state.
When $m \neq 0$,  it is an {\it ordered} state; further if all individuals have identical opinion (which is  perhaps only ideological
as opinions are continuously distributed), it is a {\it consensus} state. 
Hence consensus is not merely an agreement in this terminology.
%If there are only two surviving opinions in the system, one can say there is a polarisation. 
%other hand, if the average opinion has a nonzero value, it will be called an 
%ordered state. If in addition, it is delta-function distributed at 
%a nonzero value, it will be called a consensus state too. The latter case 
%is perhaps only ideological. 
 Obviously, in
the conserved system, the nomenclature of order and disorder is irrelevant.
We intend to locate the regions of neutrality, order and disorder in the present model
in the $\delta-\lambda$ plane.
%. and consensus state is what we call the neutral state.

%As discussed above, the result of the 
%Deffuant model is reproducable in the LCCC model and hence it is 
%more useful to adopt the idea of bounded confidence in the LCCC model
%and see the consequences.    

%In terms of consensus formation, one may argue that  for $\lambda < \lambda_c$, one indeed has consensus formation,
%where all opinions converge to the average (zero) and there is no fluctuation. However, since continuous opinions may be 
%associated usually with a rating or liking, we would like to call this state a case of no decision (e.g., in terms of 
%popularity of a film, this means neither a hit nor a flop but an average rating). On the other hand for $\lambda> lambda_c$, 
%there is a definite decision made by the population, going either in the negative or positive directions. 

\section{Results and numerical analysis} 

We take a population  of $N$ agents and let it evolve according to the dynamical rules defined above 
(i.e., eq \ref{eq:lccc} with a bounded confidence level $\delta$).

\subsection{Order parameter}

\begin{figure}
\rotatebox{0}{\resizebox*{6.5cm}{!}{\includegraphics{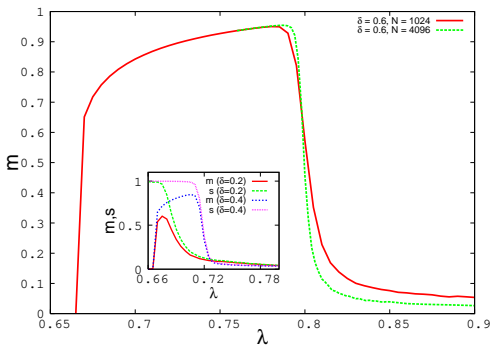}}}
\caption{(Color online)
Variation of the order parameter $m$ with $\lambda$  
for  $\delta =  0.6$ for $N = 1048$ and $4096$.  
%For $\delta = 0.2$, $s$ (Eq. (\ref{def-s}))  has also been drawn 
%for comparison. 
Inset shows $m$ and $s$ for $\delta = 0.2$ and $0.4$  against $\lambda$ for $N = 1048$.}
\label{mag-lambda}
\end{figure}

The behaviour of the order parameter after the 
system reaches equilibrium is presented in Fig. \ref{mag-lambda}.  As a function of $\lambda$, we find that the order parameter
first assumes non-zero values at a threshold value of $\lambda= \lambda_{c_1}$ which is independent of $\delta$; $\lambda_{c_1} \simeq 2/3$ as in the original LCCC model.
The order parameter increases with $\lambda$ beyond $\lambda_{c_1}$ up to a certain value of $\lambda$ and decreases to zero 
as $\lambda$ is  increased further. The decrease becomes 
steeper  with  $\delta$  and  more so when  the system size is increased. The results indicate that there are three 
distinct regions:  one ordered region for intermediate values of $\lambda$ 
and two  regions at low and 
high values of $\lambda$ where the order parameter vanishes. These two regions
may be either disordered or neutral.

We also study the behaviour of another quantity $s$ defined as 
\begin{equation}
s = \langle |f_{+} - f_{-}| \rangle,
\label{def-s}
\end{equation}
 where $f_{+}$
denotes the fraction  of population with  opinion greater than zero and $f_{-} = 1-f_+$
in a particular configuration. $\langle \cdots \rangle$ denotes average over all configurations.
It can be easily
 seen that  $s$ is equal to unity both in the neutral
state and ordered state  of LCCC (in the numerical study, we take an opinion value to be equal to zero when 
its absolute value is less than $10^{-8}$ \cite{notes}).
Deviation of $s$ from unity will indicate that opinions with both signs are present in general.
We notice that $s$  remains close to unity as $\lambda$ is increased
from zero before showing   a   sharp fall close to  a value of $\lambda$ where the order
parameter also starts to fall (Fig. \ref{mag-lambda} inset).
Evidently, as the system enters the disordered state, individual opinions are $> 0$ and $\leq 0$
with equal probability. $s$ has a monotonic behaviour  and is useful to locate the
transition to the disordered state. Moreover,
comparison of $s$ and $m$ shows that these two measures become  closer and tend to  merge
  as $\lambda$ is increased further.
 This indicates that as one moves deeper into the
disordered region, opinions become more and more close to the extreme values $\pm1$, leading to a   polarisation tendency in the opinions.

\subsection{Individual opinion distribution}

\begin{figure}
\rotatebox{0}{\resizebox*{6cm}{!}{\includegraphics{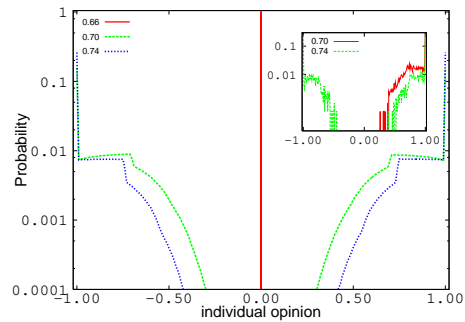}}}
\caption
{(Color online) The distribution of individual opinions shown for $N=2048$ different values of 
$\lambda $ for $\delta = 0.4$ averaged over different configurations. Inset shows the same for a single 
realisation of the system. 
For $\lambda < \lambda_{c_1} \simeq 2/3$, the average distribution  is a 
delta function at zero.
For $\lambda _{c_1} < \lambda < \lambda_{c_2} \simeq  0.7183$ (for $\delta = 0.4$), 
for a single configuration, opinions are all of one particular sign, while for averages over all configuration, the distribution is symmetric. For $\lambda > \lambda_{c_2}$, the distribution is symmetric 
even for the  single configuration.}
\label{mag-dist}
\end{figure}

To understand the nature of the phases,  
  the
distribution of individual opinions may be studied. Such studies are known to lead to a correct speculation about  phase transitions \cite{bcs2012}. This study
 shows (Fig. \ref{mag-dist}) that the probability for zero  opinion  is nearly equal to 1 below $\lambda_{c_1} \simeq 2/3$ as in the LCCC model for all $\delta$. 
%There is a small  probability ($\sim O(10^{-4})$)  for  non-zero opinions close to the extreme values and not shown in Fig. \ref{mag-dist}. 
%This probability  vanishes as system size is increased.
Hence a neutral state exists here as well and the confidence level is absolutely
irrelevant as $\lambda_{c{_1}}$ is  independent of $\delta$.

\begin{figure}
\rotatebox{0}{\resizebox*{6cm}{!}{\includegraphics{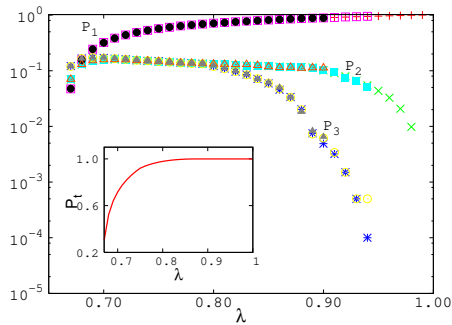}}}
\caption
{(Color online) The  probabilities $P_1, P_2$ and $P_3$ are shown against $\lambda$ for subsets A, B and C. The data for for the different subsets
 coincide 
for each of the three probabilities. Inset shows the total probability that starting with a opinion with absolute value  greater than 0.75,
it remains so at subsequent times. Data is for $\delta = 0.4$ and $N = 2048$.} 
\label{plateau}
\end{figure}

It may be noted that a plateau like region exists in the distribution where the individual 
opinions have more or less the same probability. The extent of the plateau depends on $\lambda$ and shrinks with increasing $\lambda$. 
One can investigate the reason for the existence of such a plateau region by some further analysis.
%If it happens that opinions within the interval over which the plateau extends remain within this interval with 
%equal probability,
%the  plateau can be explained.
%To see the evolution of individual opinions, one has to tag individuals. It may happen that the fluctuations are small and opinions reach more
%or less a fixed point - this is a case of  non ergodicity  as the entire opinion space is no longer accessible.
Let us take specifically the example of $\lambda = 0.74$ and  $\delta = 0.4$ where the plateau extends from about $ |o| = 0.75$ to nearly $ |o| =1$. 
It should be  noted that even in equilibrium the individual opinions undergo dynamics.
We tag  individuals with absolute value of opinion greater than 0.75 at a  time when the system has reached equilibrium and note their evolution.
First, the probability $P_t$ that these remain greater than $0.75$  in subsequent times is calculated. 
For a ergodic case, i.e., if the opinions can span the entire opinion space with equal probability, 
this probability should be around 0.25. We find, however, that $P_t$   is nearly 0.9 for 
$\lambda = 0.74$ and $\delta = 0.4$. This means that the opinions remain within this interval with a high probability and one need not really bother
about opinions with lesser values to analyse the situation in the plateau region.  $P_t$ may be close to unity   if opinions reach a 
 fixed point (or weakly fluctuate about one) and does not
imply that a plateau like region will emerge. Hence we probe further by splitting the interval into further subdivisions.
We estimate  separately $P_1$, $P_2$ and $P_3$ where, \\
(i) $P_1$ is the probability that opinions at later times are greater than 0.99\\
(ii) $P_2$ is the probability that opinions at later times lie between  0.89 and 0.99\\
(iii) $P_3$ is the probability that opinions at later times lie between  0.79 and 0.89.\\
One can define other probabilities but  these three  are sufficient for our purpose in the present case.
We also take three mutually exclusive subsets of  the tagged agents   with   initial values of $|o|$ 
between  0.75 to 0.80 (subset A),  between 0.85 to 0.9 (subset B) and  between 
0.95 to 1.0 (subset C).
Here initial value implies the value  at the time of tagging after the system
has equilibriated. 
We separate out $P_1$ as the probability of individual opinions having values  very close to unity
is much greater than the plateau height  (this is simply because of the imposed  boundary condition on the opinions) as shown in Fig. \ref{mag-dist}.  
The idea is that if opinions do reach (nearly) fixed point values,  $P_2$ and $P_3$ should be grossly different for  all the subsets A, B and C (e.g., for A, 
$P_2$ will be less than $P_3$ etc.) 
On the other hand, if  $P_2$ and $P_3$ turn out to be close to each other for any subset, one can conclude that the subspace $ 0.79 < |o| < 1.0$ 
is uniformly  accessible to the agents within that subset. Moreover if these probabilities are independent of  subset  A, B and C, 
one can argue that a  plateau region will exist.  We find   all the three probabilities $P_1$, $P_2$ and $P_3$ to be  
independent of   the subset considered (Fig. \ref{plateau}). 
 $P_1$ turns out to be  much higher than the other probabilities  as expected and indeed,  $P_2$ and $P_3$ are very close to each other for  
  $\lambda = 0.74$.  
%This immediately tells us that if the final opinion is less than $\pm 1$, its absolute value 
% can lie anywhere within $\sim  0.79$  to $0.99$ 
%with equal probability in which case a plateau is expected. 
Hence the  plateau exists and the reason is  the opinions here evolve in such a way that 
only part of the opinion space is spanned but spanned with equal probability. 
We have in fact obtained $P_1, P_2, P_3$ and $P_t$ for other values of $\lambda$  
for $\delta = 0.4$   
and shown them in Fig. \ref{plateau}.

As $\lambda$ is increased beyond $\lambda_{c_1}$,  in a single configuration, only all positive or all negative values are obtained as $N \to \infty$, while the 
average over all configurations is symmetric about zero as expected.
However, as $\lambda$ is increased further, 
%above $\lambda_{c_2}$, 
the opinions, even in a single configuration, assume both negative  
and positive values symmetrically (Fig. \ref{mag-dist} inset).
These results are consistent with the results for the quantity $s$ 
presented earlier.
Hence we infer that an order-disorder transition is taking place 
at a  value $\lambda_{c_2} > \lambda_{c_1}$ which is later confirmed 
from more detailed analysis.
% Hence we identify $\lambda_{c_2}$ to be the point of transition to a disordered state. 
For $\delta = 1$, the LCCC model, $\lambda_{c_2}$ is equal to 1 as expected.

\subsection{Phase transitions and critical properties}

The ordered, disordered and neutral regions  may be identified in a 
phase diagram in the  $\delta -\lambda$ plane. 
To obtain the phase boundaries in this  plane, 
we estimate the phase transition points by traditional methods, i.e., 
attempt finite size scaling for the relevant physical  variables, if possible. 
Among these variables is the fourth  order  Binder cumulant (BC) defined as
\begin{equation}
U = 1-\frac{\langle m^4\rangle} {3\langle m^2\rangle^2}.
\end{equation}
Here we discuss the case for $\delta < 0.3$ and $\delta \geq 0.3$ separately for reasons which will be 
clear later.

\begin{figure}
\rotatebox{0}{\resizebox*{7cm}{!}{\includegraphics{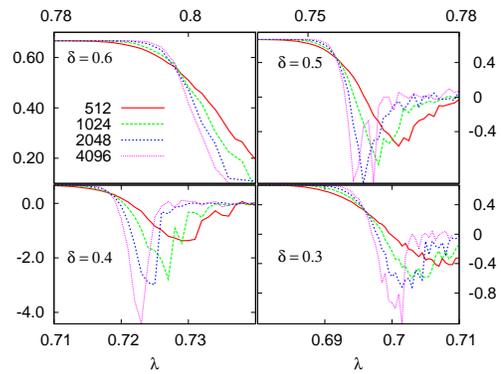}}}
\caption{(Color online) The Binder cumulant is shown for different values of $\delta$ for $N =
512, 1024, 2048$ and 4096. 
For smaller values of $\lambda$, it shows clearly that a negative dip exists which approaches the intersection point as $N$ increases. 
Other analysis suggests that the negative dip will be observed for all values of $\lambda > 0.3$ if sufficiently larger system sizes are considered (see text). 
Colour code is same for all the figures.}
\label{bc}
\end{figure}

\begin{figure}
\rotatebox{0}{\resizebox*{8cm}{!}{\includegraphics{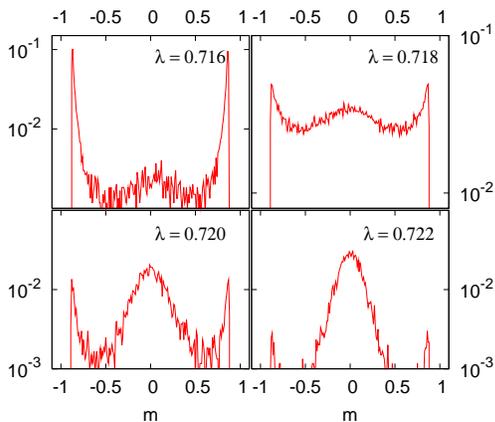}}}
\caption{(Color online) Distribution of the order parameter  close to the transition point $\lambda_{c_2} \simeq 0.7183  $ for $\delta = 0.4$ ($N = 2048$). }
\label{op-dist}
\end{figure}

$\delta \geq 0.3$:
Plotting $U$ against $\lambda$, we find that there is indeed a crossing point but interestingly, the 
BC shows a negative dip (Fig. \ref{bc}) for $\delta < 0.6$ for the system sizes considered. In fact it becomes more negative as the system size $N$ is increased and the location of 
the negative dip approaches the crossing point as well. 
These are typical indications of a first order phase transition \cite{binder}. To confirm whether a first order transition is really taking place, we also plot the distribution of 
the order parameter very close to $\lambda_{c_2}$. One expects the distribution to have peaks at nonzero values of $m$ 
below the critical point (usually the distribution is a double gaussian). 
For a continuous phase transition, as the critical point is approached from below, the peaks occur at smaller and smaller values of $m$, finally merging at $m=0$ continuously at the critical
point. 
%Above the critical point, the distribution becomes more more skewed as one goes away from the critical point. 
For a first order phase transition, on the other hand, the peaks at nonzero values of the order parameter 
remain at more or less the  same positions up to the transition point \cite{binder,binder2,challa,lee,dohm}. Here we find exactly  this behaviour 
(Fig. \ref{op-dist}); note that for finite systems,  weak peaks will still show at  nonzero values of $m$ just above the 
transition point (instead of a perfect gaussian with mean zero).

We attempt to obtain scaling forms for the BC ($U$), order parameter ($m$) and a quantity 
analogous to susceptibility per spin (in magnetic systems) given by $\chi = \frac{1}{N}[\langle M^2\rangle - \langle M \rangle ^2]$ where 
$M$ is the total opinion, $M = |\sum o_i|$.
The expected behaviour are given by 
\begin{eqnarray}
U &=&f_1((\lambda - \lambda_c{_2})N^\mu) \nonumber \\
m &=& N^{-a}f_2((\lambda - \lambda_c{_2})N^\mu) \nonumber \\
\chi &=&N^b f_3((\lambda - \lambda_c{_2})N^\mu).
\end{eqnarray}

\begin{figure}
\rotatebox{0}{\resizebox*{6cm}{!}{\includegraphics{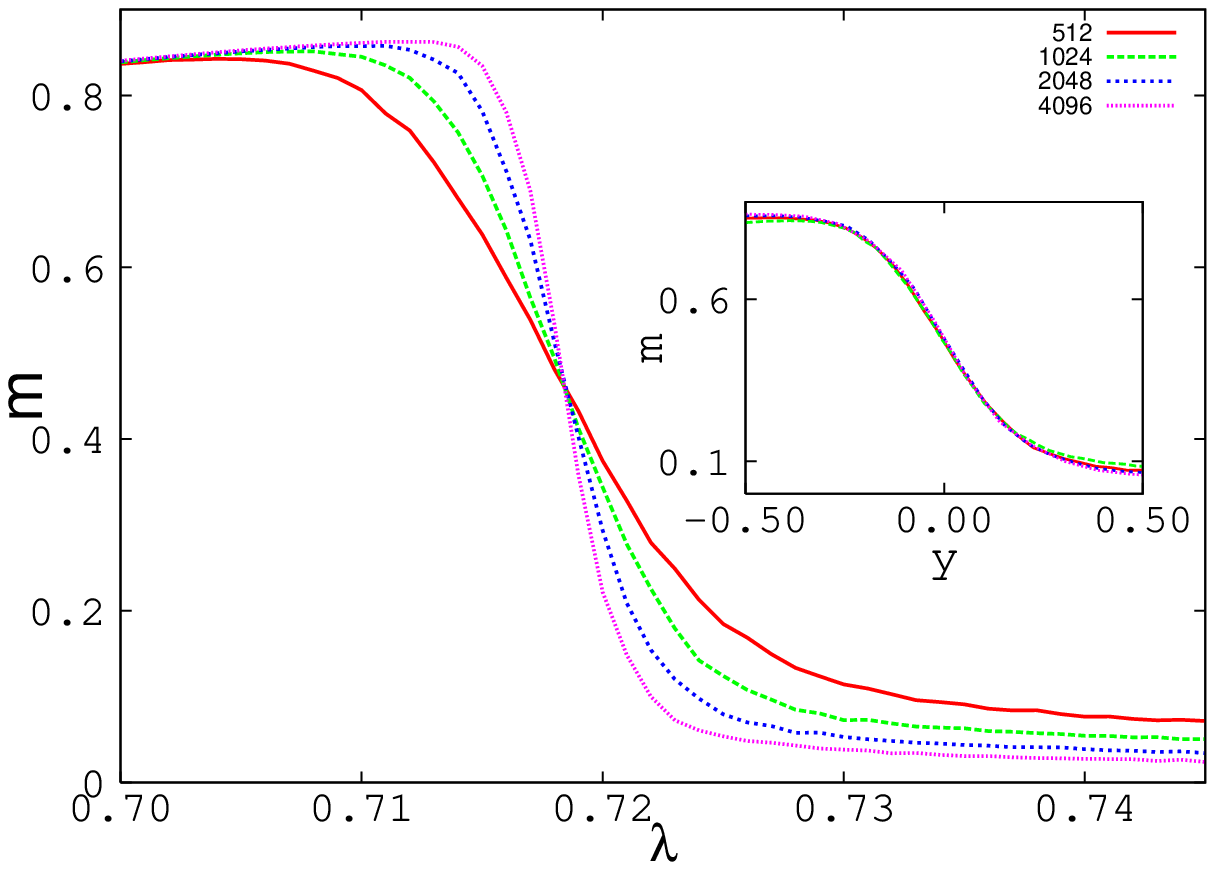}}}
\rotatebox{0}{\resizebox*{6cm}{!}{\includegraphics{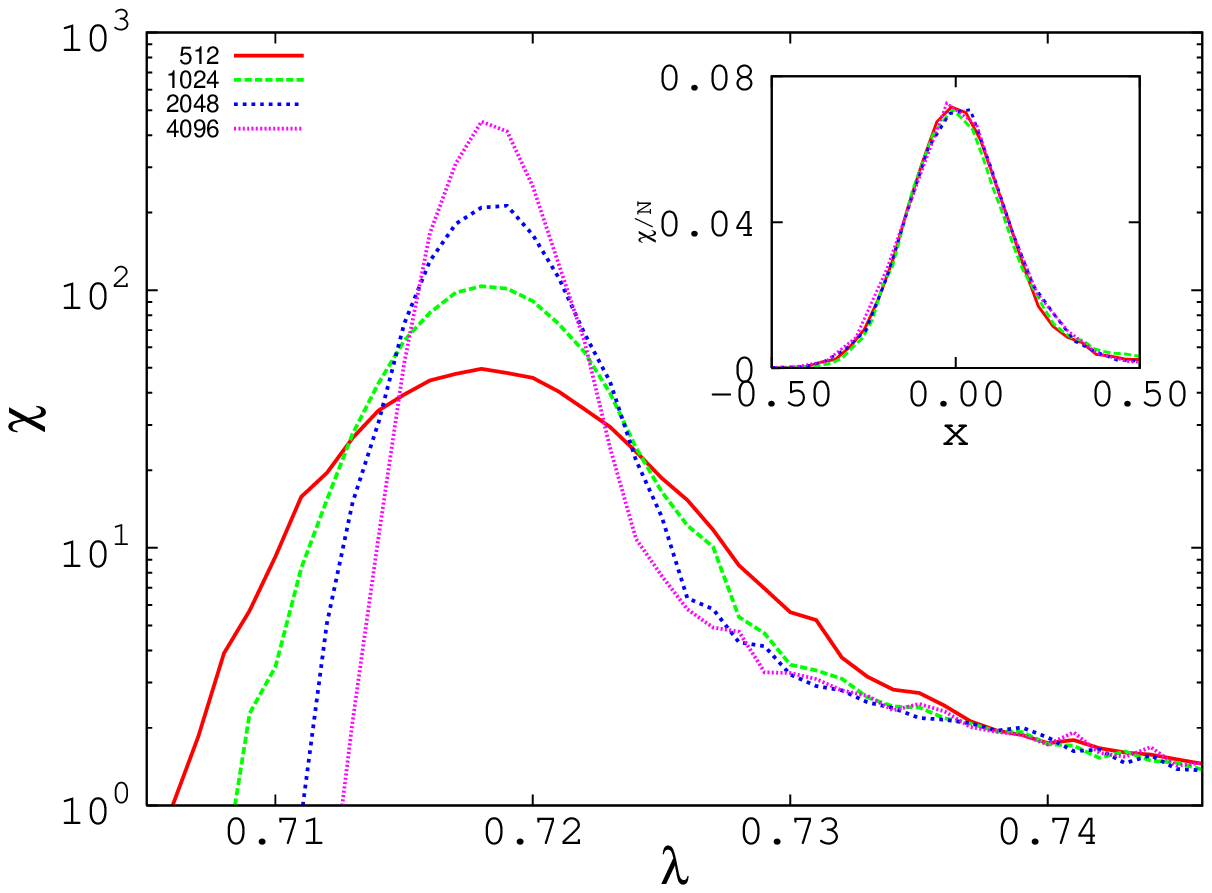}}}
\caption{(Color online) 
The variation of $m$ and $\chi$ are shown for different system sizes for $\delta = 0.4$. The insets show the scaling plots with $m$ and $\chi/N$ plotted against $x= (\lambda-0.7183)N^{0.53}$.} 
\label{collapses}
\end{figure}

For  first order phase transitions  in finite systems, 
one expects that instead of a delta function behaviour at the transition point, there will be a peak 
in the susceptibility   which will diverge with the system size. 
The order parameter exponent $a$ is expected to be close to zero. 
We find that the above scaling forms are indeed appropriate in the present case, the data collapse to 
a single curve for specific values of $a$, $b$ and $\mu$ (shown for $m$ and $\chi$ in Fig. \ref{collapses}). The value of $a$ is indeed very close to zero and 
$b \simeq 1$ for all values of $\delta \geq 0.3$. However, the value of $\mu$ appears to have a systematic variation with $\delta$. 
%Usually, one finds the scaling in first order phase transition to be dictated by ${\cal L}^d$ in a system of linear dimension ${\cal L}$ and
%dimension $d$, in the present case we do not have lattice so to say and the exponents $b$ and $\mu$  cannot be predicted.
%But the fact that $\mu$ has different values implies the first order phase transition is not universal.  
Since $\delta$ effectively puts a restriction on the number of compatible neighbours, it is not surprising that $\mu$, which is associated with $N$, 
 shows a dependence on $\delta$.
The values of the transition point $\lambda_{c_2}$ and the exponents are presented in Table \ref{table1}.

\begin{center}
\begin{table}
\caption{The transition point and  the values of the exponents 
for $\lambda > 0.3$. The typical errors in the data are $\pm 1.0 \times 10^{-4}$ for $\lambda_{c_2}$;
$\pm 0.01 $ for $\mu$, $O(10^{-3})$  for  $a$ and $O(10^{-2})$ for $b$.}
\begin{tabular}{|c|c|c|c|c|}
\hline
$\delta$ & $\lambda_{c_2}$& $\mu$ & $a$ & $b$\\
\hline
0.30 &  0.6958 & 0.56 &  0.00 &    1.04\\
0.40&   0.7183 & 0.53 &  0.00 &    1.05\\
0.50&   0.7555&   0.50&   0.00 &    1.02\\
0.60&   0.7980&   0.43&   0.00 &    0.90\\
0.70&   0.8415&   0.34&  0.035 &   0.90\\
0.80&   0.8850&    0.26&   0.00 &    1.00\\
0.90&   0.9530&    0.20&   0.025&   0.90\\
\hline
\end{tabular}
\label{table1}
\end{table}
\end{center}

All the above discussions are however, valid for $\delta > 0.3$ only. The first order phase transition is most strongly observed close to $\delta = 0.4$. As for the negative dip, it is not observed for $\delta \geq 0.6$ with $N \leq 4096$,  but the values of the exponents indicate that   
the transition is first order-like. The negative dip for $\delta \geq 0.6$ is thus expected to be observed for
even higher values of $N$ \cite{binder}.
% indicating  that perhaps the first order transition  becomes weaker as $\delta$ is increased. 
%When $\delta $ is decreased below $0.4$, the negative dip
%can be observed clearly up to $\delta = 0.3$ but the transition seems to be less strongly first order compared to that at 0.4. 

\begin{figure}
\rotatebox{0}{\resizebox*{7cm}{!}{\includegraphics{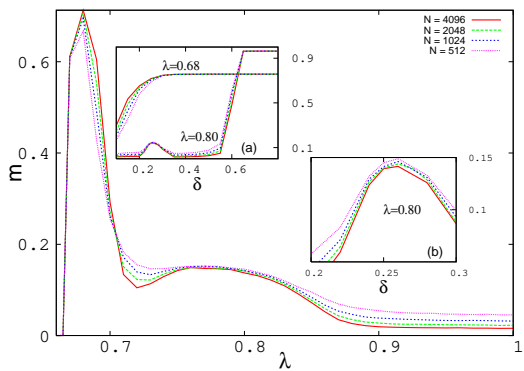}}}
\caption{(Color online) Variation of the order parameter $m$ with $\lambda$  for $\delta = 0.25$s. 
Inset (a) shows the variation of $m$ with $\delta$ for two fixed values of $\lambda$. Inset
(b)  shows a magnified portion of inset (a) to show the finite size effects.
All data are shown for four different values of $N$ following the same colour code.
} 
\label{mag-lamda-del}
\end{figure}

$\delta < 0.3$: 
When $\delta$ is decreased below 0.3,
the results do not give any clear indication about the nature of the 
phase transition and shows some anomalous behaviour.
A rather uncharacteristic behaviour of the order parameter and the Binder cumulant is observed.
The order parameter $m$ shows a nonmonotonic behaviour when 
 plotted as a function of $\delta$ with fixed $\lambda$ or vice versa (Fig. \ref{mag-lamda-del}). 
%We notice that for  e.g., 
%$\lambda = 0.70$, $m$ increases as $\delta$ increases
%up to  0.25, decreases for $0.25 <  \delta  < 0.3$ before increasing  monotonically for larger values of $\delta$ (for $N = 4096$).  
%This indicates a re-entrant behaviour of the ordered phase as $\delta$ is varied.  
%Based on the behaviour of the $m$ as system size increases,
%we estimate the transition points to the disordered region, which is a somewhat approximate method and does not
%determine the nature of the phase transition. 
A hump appears in the $m$ versus $\lambda$ plots for $\delta < 0.3$ and $\lambda \geq 0.7$ showing the existence of a local maximum value. A closer examination 
 reveals that this hump shrinks in size, albeit very slowly, when $N$ is increased for $\lambda > 0.71$ (see inset (b) of Fig. \ref{mag-lamda-del}) and is expected
to disappear in the thermodynamic limit. 
The same is true for the hump appearing in the plot of $m$ versus $\lambda$ for $\delta = 0.25$.
%suggesting a weak re-entrant behaviour for a very narrow range of $\lambda$ close to $\lambda = 0.7$.  
For $\delta < 0.3$, there are large fluctuations and irregularities  in the BC as well which does not allow one to 
do a finite size scaling analysis and get the exponents. 
%However, the BC remains positive at least for the system sizes studied here 
%($N \leq 8192$). 
The irregular behaviour
of the BC and the order parameter may be because  of the fact that the interactions become less likely to occur  as 
the confidence level is decreased. For this reason we restrict our study to $\delta \geq 0.1$. 
%In the inset of Fig. \ref{mag-lamda-del}, the $m$ versus $\lambda$ plot consistently shows a hump occurring at large 
%values of $\lambda$. This is most prominently observed close to $\delta \sim 0.25$ where the 
%local maxima of the order parameter appear in the $m$ versus $\delta$ plots.   This, however,   turns out to be a finite size behaviour 
%as expected. 

\begin{figure}
\rotatebox{270}{\resizebox*{5cm}{!}{\includegraphics{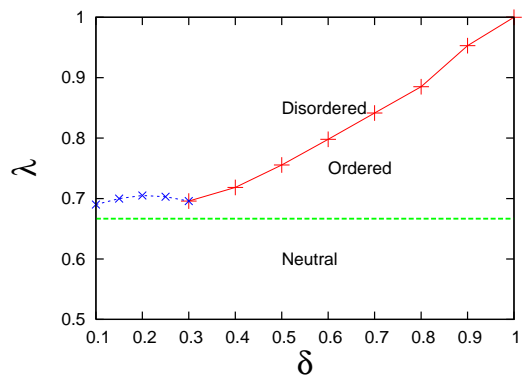}}}
\caption{(Color online) The phase diagram in the $\delta-\lambda$ plane shows the existence of the neutral region (for $\lambda \leq \lambda_c{_1} \simeq 2/3$), the ordered region and the disordered region. The ordered and disordered regions are separated by a first order boundary  (continuous line in red) for $\delta \geq 0.3$ obtained 
using a finite size scaling analysis. 
For $\delta < 0.3$, the phase boundary (broken line in blue) has been obtained approximately only from the behaviour of the order parameter (see text).} 
\label{phase-diag}
\end{figure}

We have estimated, somewhat approximately, $\lambda_{c_2}$ for $\delta < 0.3$
from the crossing point of the order parameter curves for different sizes.
For example, from the data of $m$ against $\lambda$ for $\delta = 0.25$ shown in Fig.  \ref{mag-lamda-del},
we get $\lambda_{c_2} \approx 0.7$. 
The complete phase diagram in the $\delta- \lambda$ plane is shown in Fig. \ref{phase-diag}.
% where
% the phase boundary for $\delta < 0.3$ and the  re-entrant behaviour are shown  schematically.  
%The nature of the phase transition, if any, is not apparent (it can still be first order but one has to check for larger sizes).  
%One interesting point to be noted is, in the ordered region beyond $\delta = 0.3$, along any  line $\lambda = \rm{constant}$, the value of the order parameter is 
%independent of $\delta$ which is not true for  $\delta < 0.3$. In fact, for the same value of $\lambda$, the order parameter is 
%much larger in    $\delta > 0.3$  compared to  that in $\delta < 0.3$. For example, for $\lambda$ = 0.8, $m \simeq 0.8$ for the ordered
%phase in the $\delta > 0.3$ region while for $\delta < 0.3$, the maximum value of $m$ is $\sim 0.14$. 
%Hence for  $\delta < 0.3$,  one gets a rather  weakly ordered region. Previously we had commented that in the 
%ordered region, when $\delta > 0.3$, opinions of only one sign prevail (in a single realisation) with probability equal to unity as $N \to \infty$, but for finite values
%of $N$, with very small probability, opinions of both signs may be present. 
%In the weakly
%ordered region below $\delta < 0.3$ also, opinions of one sign dominate,  however, the probability that opinions of both signs are present is much larger in comparison
%for the same value of $N$. 

\subsection{Discussions}

Let us  try to understand the results by analysing the role of the confidence level  $\delta$.
Let the $i$th  agent with opinion $o_i$ interact with another agent with opinion $o_i + x$ where $|x| \leq 2\delta$.
Then,
\begin{equation}
o_i(t+1) = \lambda (1+ \epsilon )o_i(t)  + \lambda \epsilon x.
\end{equation}

Consider the case when $\delta $ is small. 
On an average, when $\lambda$ is smaller than 2/3, the first term will
make $o_i$ smaller in magnitude while the contribution of the second term 
can be neglected. Then we find that $o_i (t) \to 0$ as $t\to \infty$ 
implying the convergence to a neutral state. 
Since for $\delta =1$, it is already known that there is a  transition to the neutral state at $\lambda \simeq 2/3$, 
we conclude that for any $\delta$ this is the case as is confirmed  by the numerical results.
Actually, when individual opinions decrease towards zero because of the effect of the first term,  the difference 
in opinions automatically becomes  less ($x \to 0$) so that large and small $\delta$ values have the same effect;
the second term does not contribute eventually.
Thus $\lambda_{c_1}$ is independent of $\delta$.
%Although  $m$ vanishes for $\lambda <  2/3$ for all $\delta$, there are subtle differences, e.g.,  
%for finite sizes, $m$ is order of magnitude higher when $\delta$ is small.  

What happens at higher values of $\lambda$? First consider    small 
values of $\delta$ again which  means $x$ is small too.
Hence the second term still contributes less compared to the first and 
$o_i$   will retain its original sign in most cases if $\lambda$ is sufficiently large. So it is expected that there will be a region where opinions of both
signs are present and $m=0$, as originally opinions are uniformly distributed with positive and negative signs.

%Since we had, to begin with, opinions of both 

%signs, for small $\delta$, there will occur opinions of both signs 

%for even not very large values of $\lambda$  after repeated interactions. 
However, if $\delta$ is large,
there is no guarantee that the second term is small and opinions will retain their original signs, unless  $\lambda$ is   also very  high. This explains
why we observe the transition to the disordered state at a higher value of  $\lambda$ as  $\delta$ is increased. 
 Also, 
 it is not surprising that there will be a  ordered region between $\lambda_{c_1} $ and $\lambda_{c_2}$ (as already it is known to be present  for $\delta = 1.0$) where the
LCCC property of all opinions assuming the same sign is still valid.

Although we have restricted to $\delta \geq 0.1$ in the numerical simulations,  the case when $\delta$ is exactly equal to zero can 
be discussed theoretically to some extent. 
%It has already been mentioned that there will be infinitesimally small number of 
%interactions. 
If an interaction takes place, 
the  opinion for the $i$th agent  follows the evolution equation
\begin{equation}
o_i(t+1) = \lambda  (1 + \epsilon) o_i(t).  
\end{equation}
 This equation is nothing but the dynamical 
equation obtained for the LCCC model in the limit of a single parameter map \cite{bkcop,asc,soumya-map},
 where  the transition to an ordered state occurs at a value of $\lambda = e/4$.
However, in the present model the above is valid only when there is a second agent  
with opinion equal to $o_i$ as well.   Since this will be  extremely rare,  effectively  most of the opinions remain frozen and the
single parameter map is not representative of all the agents' opinion evolution.  In fact, $\delta=0$ may be regarded as a 
 line of discontinuity in the phase diagram of the model in the $\delta-\lambda$ plane as it will neither  have the neutral state nor the ordered state anywhere. 
The disordered state  is also different in nature for $\delta=0$; here the individual
opinion distribution will be flat while for $\delta \neq 0$, however small, it is not so.

\section{Summary and conclusions}

In summary, we have studied a model of continuous opinion dynamics with an attempt to merge the 
concepts of confidence level and conviction. We find the interesting result that with a large value of  conviction and  
with any finite bound on the confidence level (i.e., $\delta < 1$), a disordered state exists with 
tendency to polarisation. This is indeed justified, if agents are convinced to a large extent in their
opinions and interact with like minded people only, the sign of the opinion, (representing liking/disliking) is likely 
 to be maintained,  giving rise to a polarised society. The neutral state with all opinions equal to 
zero  remains unperturbed with the introduction of the bounded confidence. 
It is found that 
at least for $\delta > 0.3$, 
the order-disorder transition is first order in nature.  
 For smaller confidence level, when $\delta < 0.3$,  
the ordered phase shrinks to a narrow region of the phase diagram.
%rue that a disordered phase exists, possibly when $\lambda_{c_2} > 0.7$.
%But, for definite combinations of values  of $\delta$ and 
%$\lambda$, there is indication that a  weak order survives in a narrow region of
%the phase diagram.
%However, this might as well be a finite size effect.

 In conclusion, we obtain   a  phase diagram with many features when the 
concept of bounded confidence is incorporated in the LCCC model of opinion dynamics. The overall result is that   
when bounded confidence level  is large, there will be order in the society provided people are 
not too rigid. In the original DNAW model also, it had been shown that above a certain confidence level,
society behaves more homogeneously.  We show that this tendency remains  true but only up to a certain level of 
conviction. This seems to be a realistic scenario  and thus the combination of concepts from two different  models in the 
present model of opinion dynamics is successful in reproducing this 
 desired feature of a society.

Acknowledgment: The author is grateful to Bikas K. Chakrabarti and Soumyajyoti Biswas for discussions and  a critical reading of the manuscript and to 
UPE projet (UGC) for financial and computational support.


\begin{thebibliography}{99}
\bibitem{notes}   Numerically, the opinions  should assume
  infinitesimally small values with both plus and minus signs
in the neutral state giving $s=0$. As
these values tend to  zero in the thermodynamic limit
 $s$ should actually be unity below $\lambda_{c_1}$ and therefore   opinion values wih aboslute value 
less than a small quantity (here $10^{-8}$) are taken as zero. 
%In the present  numerical study,  opinion $o$  has been taken to be
% zero if   $|o| < 10^{-8}$,
% which gives $s \simeq 1$  as
%shown in Fig. \ref{mag-lambda}.
For finite systems, it is also true that  opinions
have both plus and minus signs only up to a finite time
after which a 
slight imbalance in the opinion signs drives the system to a state where opinions  assume only one sign. This
occurs at longer and longer times as the system size is increased
indicating that in the thermodynamic limit, the opinions will
remain of both signs as expected.
\end{thebibliography}
\end{document}